\documentclass[cits]{PoS}
\usepackage{graphicx}
\usepackage{amssymb}
\usepackage{amsmath}

\newcommand{\betaL}{\beta_{\mathrm{L}}}
\newcommand{\betaLC}{\beta_{\mathrm{L}}^{~\mathrm{c}}}
\newcommand{\betaRef}{\beta_{\mathrm{L}}^{~\mathrm{ref}}}
\newcommand{\sub}[1]{{\scriptscriptstyle \mathrm{#1}}}
\title{Thermodynamic Study for Conformal Phase in Large $N_f$ Gauge Theory}

\ShortTitle{Thermodynamic Study for Conformal Phase in Large $N_f$ Gauge Theory}

\author{\speaker{Kohtaroh Miura}\\
        INFN Laboratori Nazionali di Frascati, 
	I-00044, Frascati (RM), Italy\\
        E-mail: \email{Kohtaroh.Miura@lnf.infn.it}}

\author{Maria Paola Lombardo\\
        INFN Laboratori Nazionali di Frascati, 
	I-00044, Frascati (RM), Italy\\
	Humboldt-Universit\"at zu Berlin, Institut f\"ur Physik, D-12489 Berlin, Germany\\
        E-mail: \email{Mariapaola.Lombardo@lnf.infn.it}}

\author{Elisabetta Pallante\\
        Centre for Theoretical Physics, University of Groningen,
	9747 AG, Netherlands\\
        E-mail: \email{e.pallante@rug.nl}}

\abstract{
We investigate the chiral phase transition at finite temperature ($T$)
in colour SU$(N_c=3)$ Quantum Chromodynamics (QCD)
with six species of fermions ($N_f=6$)
in the fundamental representation~\cite{Miura:2011mc}.
The simulations have been performed
by using lattice QCD with improved staggered fermions.
The critical couplings $\betaLC$ for the chiral phase transition
are observed for several temporal extensions $N_t$,
and the two-loop asymptotic scaling
of the dimensionless ratio
$T_c/\Lambda_{\mathrm{L}}$
($\Lambda_{\mathrm{L}}=$ Lattice Lambda-parameter)
is found to be achieved for $N_t\ge 6$.
Further, we collect $\betaLC$
at $N_f=0$~(quenched), and $N_f = 4$ at a fixed $N_t = 6$
as well as $N_f=8$ at $N_t = {6, 12}$,
the latter relying on our earlier study.
The results are consistent
with enhanced fermionic screening at larger $N_f$.
The ratio $T_c/\Lambda_{\mathrm{L}}$ depends very mildly 
on $N_f$ in the $N_f=0 - 4$ region, 
begins increasing at $N_f = 6$,
and significantly grows up  at $N_f = 8$,
as $N_f$ reaches to the edge of the conformal window.
We discuss the interrelation of the results with preconformal
dynamics in the light
of a functional renormalization group analysis.}
\FullConference{ The XXIX International Symposium on Lattice Field Theory - Lattice 2011\\
July 10-16, 2011\\
Squaw Valley, Lake Tahoe, California}
\begin{document}
\section{Introduction}
Emergence of a conformal symmetry
and  a preconformal (walking) behavior
in strongly flavored non-Abelian gauge theories
has received much attention.
Walking dynamics near the infra-red fixed point
has been advocated as a basis for strongly interacting
mechanisms of electroweak symmetry breaking.
Lattice Monte-Carlo simulations
are expected to provide a solid theoretical base
to understand the (pre-)conformal nature in the gauge theory.

A second zero of the two-loop
beta-function of massless QCD with $N_f$ flavours
implies, at least perturbatively,
the appearance of an infrared
fixed point (IRFP) at $N_f\gtrsim 8.05$ \cite{IR_Seminal}
with the restoration of conformal symmetry
before the loss of asymptotic freedom (LAF) at
$N_f^{\mathrm{LAF}}=16.5$.
Conformality should emerge
when the renormalized coupling at the would be IRFP
is not strong enough to break chiral symmetry.
This condition provides 
the lower bound $N_f^c$ of a so called conformal window in
the flavor space,
and we find elaborated analytic predictions~\cite{Miransky:1997,IR_Analytic}:
for instance, the functional renormalization group method~\cite{BraunGies}
suggests $N_f^c\sim 12$.
Before the emergence of conformal symmetry,
a qualitative change of dynamics
is claimed at $N_f=6$ based on
instanton study~\cite{Velkovsky:1997fe}.

Recent lattice studies\cite{DelDebbio:2010zz} 
focused on the computation of
the edge of the conformal window ${N_f}^c$
and the analysis of the conformal window itself, 
either with fundamental fermions
\cite{Appelquist:2011dp,Appelquist:Conformal,
Deuzeman:2008sc,Deuzeman:2009mh,Hasenfratz:2010fi, Hasenfratz:2009ea, 
Fodor:2011tu,Fodor:2009wk}.
or other representations \cite{Non_Fundamental}.
Among the many interesting results with fundamental fermions,
we single out the observation 
that QCD with three colours and eight flavours
is still in the hadronic phase
\cite{Appelquist:Conformal,Deuzeman:2008sc},
while $N_f = 12$ seems to be close to ${N_f}^c$,
with some groups favouring conformality
\cite{Appelquist:2011dp,Appelquist:Conformal,
Deuzeman:2009mh,Hasenfratz:2010fi}, 
and others chiral symmetry breaking~\cite{Fodor:2011tu}.
The onset of new strong dynamics at $N_f = 6$ 
has been implied via an enhancement of 
the ratio of chiral condensate to cubed pseudoscalar decay constant
\cite{Appelquist:2009ka}.

Using the thermal transition
as a tool for investigating preconformal dynamics 
has been largely inspired by a renormalization group analysis
\cite{BraunGies}.
The critical temperature
for the chiral phase transition
has been obtained as a function of $N_f$.
Then the onset of the conformal window has been estimated
by locating the vanishing critical temperature.
The phase transition line is almost linear with $N_f$ 
for small $N_f$, and clearly elucidates the universal critical
behaviour at zero and non-zero temperature in the vicinity of $N_f^c$.
Thus, it would be a promising direction
to extend the knowledge of finite $T$ lattice QCD
to the larger $N_f$ region, 
by using the FRG results as analytic guidance.

In this proceedings,
we investigate the thermal chiral phase transition
for $N_f=6$ colour SU$(N_c=3)$ QCD
by using lattice QCD Monte Carlo simulations
with improved staggered fermions based on
our recent study~\cite{Miura:2011mc}.
$N_f=6$ is expected to be in the important regime
as suggested by the results
in Refs.~\cite{Velkovsky:1997fe,Appelquist:2011dp}.
We also compute the critical couplings
for $N_f=0$ (quenched) and $N_f=4$ at $N_t = 6$,
and use the results from Ref.~\cite{Deuzeman:2008sc} for $N_f=8$. 
Then we investigate $N_f$ dependences of the chiral phase transition.

\section{Simulation setups}
Simulations have been performed
in the same as in the study used for $N_f=8$ in Ref.~\cite{Deuzeman:2008sc}:
We have utilized
the publicly available MILC code~\cite{MILC}
with the use of
an improved version of the staggered action, the
Asqtad action, with a one-loop Symanzik
\cite{Bernard:2006nj,LuscherWeisz}
and tadpole~\cite{LM1985} improved gauge action.
The tadpole factor $u_0$ is determined
by performing zero temperature simulations
on the $12^4$ lattice, and used as an input
for finite temperature simulations.

To generate configurations with mass degenerate
dynamical flavours,
we have used the rational hybrid Monte Carlo algorithm
(RHMC)~\cite{Clark:2006wq}.
Simulations for $N_f=6$ have been performed by using two
pseudo-fermions, and subsets of trajectories for
the chiral condensates and Polyakov loop
have been compared with those obtained
by using three pseudo-fermions with the same
Monte Carlo time step $d\tau$
and total time length $\tau$ of a single trajectory.
We have observed very good
agreement between the two cases
for both evolution and thermalization.
We have monitored the Metropolis acceptance and reject ratio,
and adjusted $\tau=0.2 - 0.24$ and $d\tau=0.008 - 0.016$
to realize the best performance.

Measured observables are the expectation values of the
chiral condensate and Polyakov loop,
\begin{align}
a^3\langle\bar{\psi}\psi\rangle =
\frac{N_f}{4N_s^3N_t}
\Big\langle\mathrm{Tr\bigl[M^{-1}\bigr]}\Big\rangle
\ ,\quad
L =
\frac{1}{N_cN_s^3}\sum_{\mathbf{x}}
\mathrm{Re}
\bigg\langle
\mathrm{tr}_c\prod_{t=1}^{N_t}U_{4,t\mathbf{x}}
\bigg\rangle
\ ,\label{eq:observ}
\end{align}
where $N_s~(N_t)$ represents the number of lattice sites
in the spatial (temporal) direction,
$U_{4,t\mathbf{x}}$ is the temporal link variable,
and $\mathrm{tr}_c$ denotes the trace in colour space.
The output of this measurement
is the critical coupling $\betaLC$ for
the chiral phase transition.

\section{Results}
\begin{figure}[ht]
\includegraphics[width=7.0cm]{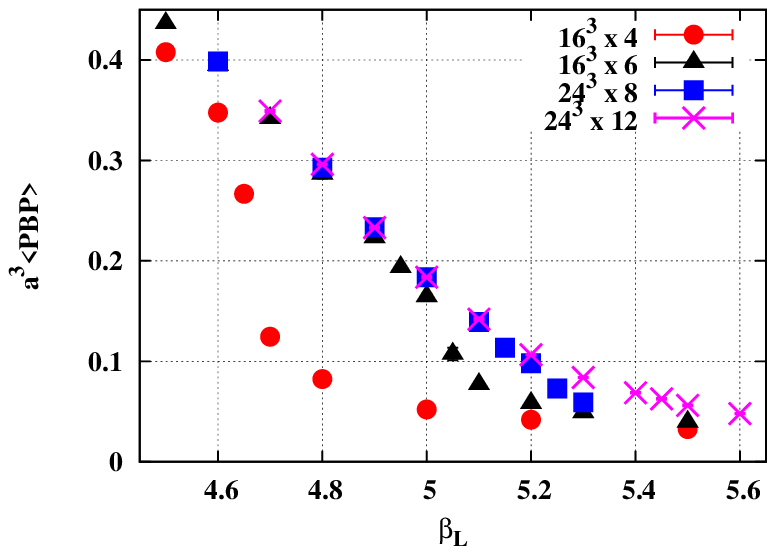}
\includegraphics[width=7.0cm]{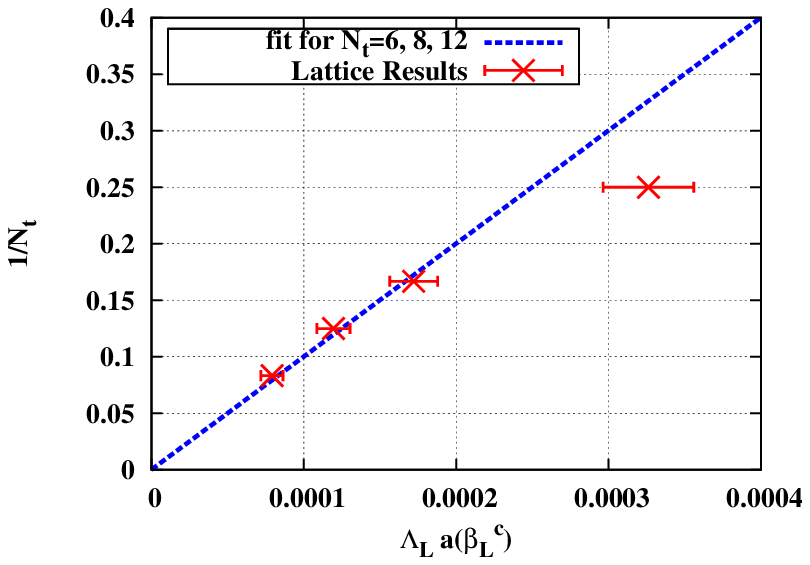}
\caption{Left:
The chiral condensate
$a^3\langle\bar{\psi}\psi\rangle$
for $N_f=6$ and $am=0.02$ in lattice units,
as a function of $\beta_{\mathrm{L}}$, for $N_t=4,~6,~8$, and $12$.
Error-bars are smaller than symbols.
Right:
The thermal scaling behaviour of the 
critical lattice coupling $\betaLC$.
Data points for $\Lambda_{\mathrm{L}}~a(\betaLC)$
at a given $1/N_t$ are obtained by using $\betaLC$
from Table~\protect\ref{Tab:bc} as input for extracting
$\Lambda_{\mathrm{L}}~a(\betaLC)$
in the two-loop expression Eq.~(\protect\ref{eq:2Loop_AS_Lat}).
The dashed line is a linear fit
with zero intercept to the data with $N_t > 4$.}
\label{Fig:Nf6}
\end{figure}

All results have been obtained for 
a fermion bare lattice mass $am=0.02$.
In the left panel of Figs.~\ref{Fig:Nf6},
the expectation values of
the chiral condensate $a^3\langle\bar{\psi}\psi\rangle$
are displayed
as a function of $\beta_{\mathrm{L}}$ for several $N_t$.
It is found that different $N_t$
give a different behaviour of $a^3\langle\bar{\psi}\psi\rangle$.
The asymptotic scaling analysis
below will confirm that it corresponds to 
a thermal chiral phase transition (or crossover)
in the continuum limit.

All values of the critical lattice coupling $\betaLC$
are summarized in Table~\ref{Tab:bc}.
For larger $N_t$, the signal for the chiral phase transition
becomes less clear, hence
we investigate the histogram of the chiral condensate:
The histogram for $N_t=8$ exhibits the double-peak structure
at $\beta_{\mathrm{L}}=5.2$, {\em i.e.},
the competition between chirally symmetric and broken vacua.
The critical coupling can be estimated as
$\betaLC=5.225(25)$ for $N_t=8$.
For $N_t=12$, we also observe the double-peak structure
in the histogram of the chiral condensate around $\beta_{\mathrm{L}}=5.45$.

These results can be analyzed and interpreted in terms of
the two-loop asymptotic scaling.
Let us consider the two-loop lattice beta function,
\begin{align}
&\beta({g})
=-(b_0 {g}^3 + b_1 {g}^5)\ ,\label{eq:beta_func}\\
&(b_0,~b_1)
=
\bigl(
(11-2N_f/3)/(4\pi)^2,\
(102-38N_f/3)/(4\pi)^4
\bigr)\ ,\label{eq:b01}
\end{align}
for fundamental fermions in colour SU$(3)$.
From  Eq.~(\ref{eq:beta_func}),
we obtain the well known two-loop asymptotic scaling,
\begin{align}
\Lambda_{\mathrm{L}}~a(\beta_{\mathrm{L}})
&=
\bigl(2N_cb_0/\betaL\bigr)^{-b_1/(2b_0^2)}
\exp\bigl[
-\betaL/(4N_cb_0)
\bigr]
\ .\label{eq:2Loop_AS_Lat}
\end{align}
Here, $\Lambda_{\mathrm{L}}$ is
the so-called lattice Lambda-parameter,
and $\beta_L = 2 N_c/ g^2$,
with $g = \sqrt{2N_c/10}\cdot g_{\mathrm{L}}$.
This definition effectively takes account of
the improvement of the staggered lattice action 
when comparing to the asymptotic scaling law, see 
Ref.~\cite{Deuzeman:2008sc}. 
We insert $\Lambda_{\mathrm{L}}$ to
the definition of temperature
$T\equiv [a(\beta_{\mathrm{L}})N_t]^{-1}$,
\begin{align}
N_t^{-1}=(T_c/\Lambda_{\mathrm{L}})
\times \Bigl(\Lambda_{\mathrm{L}}~a(\betaLC)\Bigr)
\ ,\label{eq:TcLambda}
\end{align}
and extract the physical quantity
$T_c/\Lambda_{\mathrm{L}}$
by substituting the simulation outputs $\betaLC$
for Eq.~(\ref{eq:TcLambda}).
This ratio must be unique
as long as the asymptotic scaling
Eq.~(\ref{eq:2Loop_AS_Lat})
is verified for a given $\betaLC$.

In the right panel of Fig.~\ref{Fig:Nf6},
the slope of the line connecting the origin
and the data points corresponds to  $T_c/\Lambda_{\mathrm{L}}$.
The $N_t=6,~8$, and $12$ points have
a common slope to a very good approximation,
while the $N_t=4$ result falls on a smaller slope.
The latter is interpreted as
a scaling violation effect due to the use of a too small $N_t$.
The existence of a common $T_c/\Lambda_\mathrm{L}$ for $N_t\geq 6$
indicates that the data are consistent with the two-loop asymptotic scaling
Eq.~(\ref{eq:2Loop_AS_Lat}), 
confirms the thermal nature of the transition and that $N_f=6$ is 
outside the conformal window, as expected from a previous $N_f=8$ study
\cite{Deuzeman:2008sc}.
A linear fit provides 
$T_c/\Lambda_{\mathrm{L}} = 1.02 (12) \times  10^3 $,
which can be interpreted as the value in the continuum limit
for $N_f=6$ QCD.

In order to have a more complete overview,
we have performed simulations for the theory
with $N_f=0$ (quenched) and $N_f=4$, only at 
$N_t = 6$. These theories are of course very well investigated,
however we have not found
in the literature results for the same action as ours.
We note that in a previous lattice study with improved staggered fermions 
\cite{Gupta:2000hr}, asymptotic scaling was observed
for $N_t \ge 6$ for $0 \le N_f \le 4$.
Table~\ref{Tab:bc} shows a summary of our results for 
the critical coupling  $\betaLC$
of the chiral phase transition at finite temperature 
for $N_f=0,~4,~6$, and $8$ - 
the latter from  Ref.~\cite{Deuzeman:2008sc}.

\begin{figure}[ht]
\includegraphics[width=7.0cm]{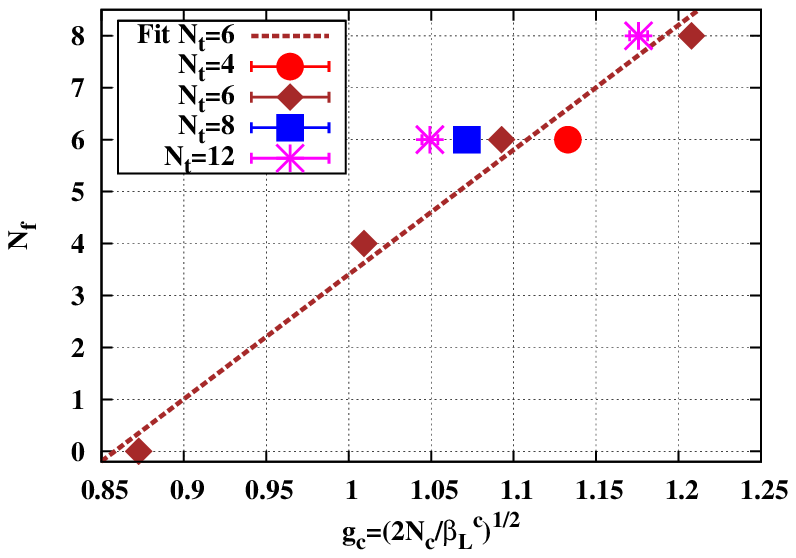}
\includegraphics[width=7.0cm]{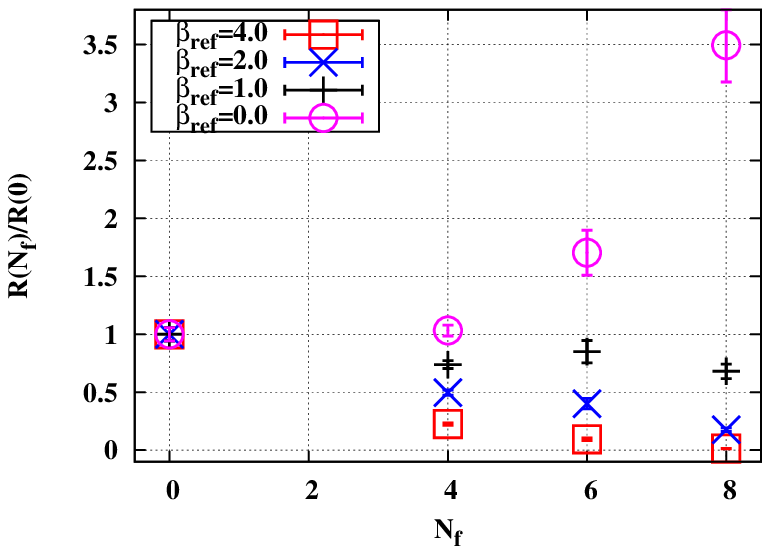}
\caption{Left:
Critical values of the lattice coupling
$g_c=\sqrt{2N_c/\betaLC}$ for theories with $N_f=0,~4,~6,~8$ 
and for several values of $N_t$ in the Miransky-Yamawaki phase diagram.
The dashed (brown) line is a linear fit to the $N_t=6$ results.
Right:
The $N_f$ dependence of
$R(N_f)/R(0)$ for several finite fixed $\betaRef$.
Here, $R(N_f)\equiv (T_c/\Lambda_{\mathrm{ref}})(N_f)$.}
\label{Fig:NfX}
\end{figure}

In the left panel of Fig.~\ref{Fig:NfX},
we display the critical values of the lattice coupling
$g_c=\sqrt{2N_c/\betaLC}$ from Table \ref{Tab:bc}
in the Miransky-Yamawaki phase diagram.
Consider the $N_t = 6$ results:
it is expected that an
increasing number of flavours favors
chiral symmetry restoration.
Indeed, we find that, on a fixed lattice,
the critical coupling increases with $ N_f$
in agreement with early studies and naive reasoning. 
The precise dependence of the critical coupling on $N_f$ at fixed $N_t$
is not known.
It is, however,
amusing to note that the results seem to be
smoothly connected by an almost straight line:
the brown line in the plot is a linear fit to the data.
Comparing the trend for $N_f = 6$ to the one for $N_f = 8$
for varying $N_t$,
one can infer a decreasing in magnitude (and small) step scaling function,
hence a walking behaviour.
Further study is needed at larger $N_f$,
and by using the same action used for $N_f=0-8$,
to confirm or disprove it.

\begin{table*}[ht]
\caption{
Summary of the critical lattice couplings $\betaLC$
for the theories with $N_f=0,~4,~6,~8$, $am=0.02$
and varying $N_t=4,~6,~8,~12$.
All results are obtained using the same lattice action.
}\label{Tab:bc}
\begin{center}
\begin{tabular}{c|c|c|c|c}
\hline
$N_f\backslash N_t$ &
$4$&
$6$&
$8$&
$12$\\
\hline
$0$ &
-&
$7.88\pm 0.05$&
-&
-\\
$4$ &
-&
$5.89\pm 0.03$&
-& 
\\
$6$ &
$4.675\pm 0.025$&
$5.025\pm 0.025$&
$5.225\pm 0.025$&
$5.45\pm 0.05$\\
$8$ &
-&
$4.1125\pm 0.0125$&
-&
$4.34\pm 0.04$\\
\hline
\end{tabular}
\end{center}
\end{table*}

Next, we study 
the $N_f$ dependence of the ratio
$T_c/\Lambda_{\mathrm{L}}$ and related quantities.
%
%
%
%
%
In addition to the scale $\Lambda_{\mathrm{L}}$,
we introduce more UV reference energy scale $\Lambda_{\mathrm{ref}}$,
which is associated with
a reference coupling $\betaRef$.
Then Eq.~(\ref{eq:2Loop_AS_Lat}) is generalized as
\begin{align}
\Lambda_{\mathrm{ref}}(\betaRef)~a(\betaL)
=
\Biggl(
\frac{b_1}{b_0^2}~
\frac{\betaL+2N_cb_1/b_0}
{\betaRef+2N_cb_1/b_0}
\Biggr)^{b_1/(2b_0^2)}\exp
\Biggl[-\frac{\betaL-\betaRef}{4N_cb_0}
\Biggr]\ .
\label{eq:int_as}
\end{align}
At leading order of perturbation theory $b_1\to 0$,
we find
$\Lambda_{\mathrm{ref}}/\Lambda_{\mathrm{L}}=\exp[\betaRef/(4N_cb_0)]$.
This equation would be analogous of the ratio
$\Lambda_{\mathrm{L}}/\Lambda_{\mathrm{MS}}$
derived in \cite{Kawai:1980ja}
for Wilson fermions up to a further linear
dependence on $N_f$ in the numerator of the exponent.
In a nutshell, the difference originates
from the fact that
we are fixing a bare reference coupling
$\betaRef$, which will be specified later.
Notice that
by construction $\Lambda_{\mathrm{ref}}$
reproduces the lattice Lambda-parameter
$\Lambda_{\mathrm{L}}$ in the limit
$\Lambda_{\mathrm{ref}}(\betaRef\to 0)=\Lambda_{\mathrm{L}}\bigl(1+\mathcal{O}(1/\betaLC)\bigr)$.

Let us consider first
$R(N_f)|_{\betaRef=0.0}=T_c/\Lambda_{\mathrm{L}}$.
The values of $T_c/\Lambda_{\mathrm{L}}$
are found to be
$600\pm 34$,
$620\pm 28$,
$1023\pm 117$, and
$2098\pm 191$
for $N_f={0,~4,~6}$, and $8$, respectively,
and represented as circles in the right panel of Fig.~\ref{Fig:NfX}
(the vertical axis
is normalized by $R(0)=(T_c/\Lambda_{\mathrm{L}})(N_f=0)$
for each $\betaRef$).
The ratio does not show a significant $N_f$ dependence
in the region $0\leq N_f\leq 4$, it  
starts increasing at $N_f = 6$,
and undergoes a rapid rise around $N_f=8$.
The nearly constant nature of $T_c/\Lambda_{\mathrm{L}}$
in the region $N_f\le 4$ indicates that
the role of such  energy scale is not significantly
changed by the variation of $N_f$ (see  \cite{Braun:2009si}
for a detailed discussion of this point.)
In turn, the increase of
$T_c/\Lambda_{\mathrm{L}}$ in the region $N_f\ge 6$
might well imply that the chiral dynamics
becomes different from the one for $N_f\leq 4$.
Indeed, a recent lattice study~\cite{Appelquist:2009ka}
indicates that $N_f = 6$ is close to
the threshold for preconformal dynamics.

We now consider $T_c/\Lambda_{\mathrm{ref}}$
with finite $\betaRef$. 
The $N_f$ dependence of the ratio 
$R(N_f)\equiv (T_c/\Lambda_{\mathrm{ref}})(N_f)$ is shown
for several $\betaRef$ in the right panel of Fig.~\ref{Fig:NfX},
(with normalization
by $R(0)=(T_c/\Lambda_{\mathrm{ref}})(N_f=0)$
for each $\betaRef$).
$T_c/\Lambda_{\mathrm{ref}}$ is now
a decreasing function of $N_f$
for a larger $\betaRef$.
The $\Lambda_{\mathrm{ref}}$
associated with a $\betaRef\gg\beta_*=2N_c/g^2_{\sub{\mathrm{IRFP}}}$
would be less sensitive to the IR or chiral dynamics.
Assuming $N_f^c\simeq 12$,
the two-loop beta-function
leads to $\beta_*=-2N_cb_1/b_0\simeq 0.63$.
The decreasing nature of
$(T_c/\Lambda_{\mathrm{ref}})(N_f)$
is found to start around $\betaRef = 1.0\gtrsim \beta_*$.
Thus, the use of a UV reference scale
leads to the decreasing $(T_c/\Lambda_{\mathrm{ref}})(N_f)$.
This trend is consistent with the FRG study~\cite{BraunGies},
where the decreasing $T_c(N_f)$
{has been obtained by using the $\tau$-lepton mass
$m_{\tau}$} as a common UV reference scale
with a common coupling $\alpha_s(m_{\tau})$.
We note that we have constrained our analyses
$\betaRef < \beta_{\mathrm{UV}}=\betaLC(N_f)\leq 4.1125\pm 0.0125$.

With the use of a UV reference scale,
we should observe the predicted critical behavior
\cite{BraunGies},
\begin{align}
T_c(N_f) = K |N_f - N_f^c|^{-1/ \theta}\ .
\end{align}
By choosing the critical exponent
$\theta$ in the range predicted by FRG:
$1.1 < 1/|\theta| < 2.5$,
our data are consistent with the values  
$N_f^c = 9(1)$ for $\betaRef = 4.0$ and $N_f^c = 11(2)$ for
$\betaRef = 2$. We plan to extend and refine this
analysis in the future, and here we only notice a reasonable
qualitative behaviour.

\section{Summary}
We have studied the chiral phase transition at finite $T$
for colour $SU(3)$ QCD with $N_f=6$
by using lattice QCD Monte-Carlo simulations
with improved staggered fermions~\cite{Miura:2011mc}.
We have determined the critical lattice coupling
$\betaLC$ for several lattice temporal extensions
$N_t$, and extracted the dimensionless ratio
$T_c/\Lambda_{\mathrm{L}}$
($\Lambda_{\mathrm{L}}=$Lattice Lambda-parameter)
by using two-loop asymptotic scaling.
The analogous result for $N_f=8$ has been extracted
from Ref.~\cite{Deuzeman:2008sc}.
$T_c/\Lambda_{\mathrm{L}}$ 
for $N_f=0$ and $N_f=4$ has been measured at fixed $N_t=6$,
barring asymptotic scaling violations.
Then we have discussed the $N_f$ dependence of the ratios
$T_c/\Lambda_{\mathrm{L}}$ and $T_c/\Lambda_{\mathrm{ref}}$,
where $\Lambda_{\mathrm{ref}}$ is a UV reference energy scale, 
related to $\Lambda_{\mathrm{L}}$ via
$\Lambda_{\mathrm{ref}}/\Lambda_{\mathrm{L}}\simeq\exp[\betaRef/(4N_cb_0)]$.
We have observed that $T_c/\Lambda_{\mathrm{L}}$
shows an increase in the region 
$N_f=6-8$, while it is approximately
constant in the region $N_f\leq 4$. 
We have discussed this qualitative change
for $N_f\geq 6$ and a possible relation with a preconformal phase.
The ratio $T_c/\Lambda_{\mathrm{ref}}$
is a decreasing function of $N_f$.
This behaviour is consistent with
the result obtained
in the functional renormalization group analysis~\cite{BraunGies}.
Next steps of the current project involve a scale setting
at zero temperature by measuring a common UV observable.

\section*{Acknowledgements}
We thank Holger Gies, Jens Braun,
Michael M\"uller-Preussker, Marc Wagner, Biagio Lucini,
Volodya Miransky, Albert Deuzeman, and Tiago Nunes da Silva
for fruitful discussions. 
This work was in part based on the MILC Collaboration's public
lattice gauge theory code~\cite{MILC}.
The numerical calculations were carried out on the
IBM-SP6 at CINECA,
Italian-Grid-Infrastructures in Italy,
and the Hitachi SR-16000 at YITP, Kyoto University in Japan.


\end{document}